\documentclass[10pt,
               aps,
               twocolumn,
               prb]
              {revtex4-2}
\usepackage{graphicx}
\usepackage{dcolumn}
\usepackage[protrusion=true]{microtype}

\usepackage{losc_macros}

\usepackage[hidelinks]{hyperref}                 

\begin{document}
\title{Correcting Delocalization Error in Materials with Localized Orbitals
       and Linear-Response Screening}
\author{Jacob Z. Williams}
\altaffiliation{Present address:~Theoretical Division, 
                Los Alamos National Laboratory, Los Alamos, NM 87545, USA}
\affiliation{Department of Chemistry, Duke University, Durham, NC 27708, USA}
\author{Weitao Yang}
\email{weitao.yang@duke.edu}
\affiliation{Department of Chemistry, Duke University, Durham, NC 27708, USA}
\affiliation{Department of Physics, Duke University, Durham, NC 27708, USA}
\date{\today}

\begin{abstract} 
Delocalization error prevents density functional theory (DFT)
from reaching its full potential,
causing problems like systematically underestimated band gaps
and misaligned energy levels at interfaces.
We introduce lrLOSC to correct delocalization error in materials
over a wide range of band gaps.
We predict eleven materials' fundamental gaps to within
\qty{0.22}{\electronvolt}, while offering a nonzero total energy correction;
molecular properties are improved with a parallel implementation
of the same theory
[\emph{J.~Phys.~Chem.~Lett.}~\href{https://doi.org/10.1021/acs.jpclett.5c00120}
                                  {\textbf{16}, 2492 (2025)}].
lrLOSC is an essential step toward modeling molecules, materials, 
and their interfaces within the same DFT framework. 
\end{abstract}

\maketitle

\section{Introduction}
Kohn--Sham density functional theory (DFT)
\cite{hohenberg1964, kohn1965, levy1979, levy1982, lieb1983} is rightly 
regarded as the default method for quantum-mechanical calculations in both
chemistry and materials science. The availability of reasonably accurate
density functional approximations (DFAs), implemented efficiently in community
software, remains unparalleled decades later. Frontier orbital---in bulk systems,
valence band maximum and conduction band minimum---energies from (generalized)
Kohn--Sham DFAs have been shown to be the chemical potentials, defined as the
derivatives of the associated DFA's total energy with respect to the electron
number \cite{cohen2008a, yang2012}; the proof uses Janak's theorem
\cite{janak1978}, but extends it to functionals of the density matrix. DFA
frontier orbitals thus predict the fundamental gaps of molecules and materials
based on the exact conditions for fractional charges
\cite{perdew1982, yang2000}. But DFAs suffer characteristic and enduring
systematic errors. Although they generally predict accurate total energies for
equilibrium structures, glaring problems appear even for simple molecules far
from equilibrium. For example, commonly used semilocal functionals are not
size-consistent: they predict an unphysically low total energy for \ce{H2+} at
the dissociation limit, with $E(\ce{H} \cdots \ce{H+}) < E(\ce{H}) + E(\ce{H+})$
\cite{zhang1998}. A significant underestimation of the fundamental (band) gap has
also been known for decades \cite{perdew1985}. In addition to the frontier
orbitals, band structures computed by (semilocal) DFAs are quite inaccurate, and
decades of work have been poured into various methods for improving them
\cite{hybertsen1986, dabo2010, tsuneda2010, kronik2012, korzdorfer2012,
      puschnig2017}.

Computing accurate band structures, and especially band gaps, within the
efficient DFT framework is a pressing problem for computational physicists,
chemists and materials scientists. Predicting and aligning energy levels is
critical for the development of semiconductor technology
\cite{heyd2005, xiao2011}; solar cells \cite{wang2018}; and photocatalysts
\cite{li2017b}. Interfacial band structure \cite{ishii1999, braun2009} is
particularly challenging because adsorbates' bands are renormalized, yet of
outsize importance in predicting the properties of next-generation heterogeneous
materials \cite{kroemer2001, editorialboard2012}. Questions of size-consistency,
while less prominent for calculations for bulk materials, resurface when
computing interfaces between finite and bulk systems. 

Behind DFAs' problems with dissociation limits, band gaps, and band structures
lies \emph{delocalization error}
\cite{mori-sanchez2008, cohen2008, cohen2012, bryenton2022}. Delocalization
error in DFAs is a systematically incorrect behavior of the
energy $E$ viewed as a function of the number $N$ of electrons, with a
manifestation that depends on the size of the system \cite{mori-sanchez2008}.
The exact $E(N)$ curve is piecewise linear, with derivative discontinuities at
integer $N$ \cite{perdew1982, yang2000}; the magnitude of the discontinuity
gives the fundamental gap. $E(N)$ is strictly convex in finite systems when
calculated with typical DFAs, underestimating the derivative discontinuity and
thus the gap \cite{cohen2008a}. As the size of the system increases, however,
$E(N)$ becomes less convex. In periodic boundary conditions, piecewise
linearity of $E(N)$ is restored automatically, but deceptively: delocalization
occurs across the entire infinite lattice, the derivative discontinuity is
underestimated just like for finite $N$, and the total energy of charged bulk
systems is predicted to be incorrect \cite{mori-sanchez2008}. Correcting
delocalization error implies computing the derivative discontinuity in $E(N)$
correctly to yield correct orbital energies. Doing so size-consistently,
however, also requires correcting the total energy of the underlying DFA.

Two ingredients---orbital localization and screening---are key to delocalization
error corrections, especially in bulk systems. In very small molecules,
delocalization error can be corrected with reasonable accuracy without
accounting for either. The global scaling correction \cite{zheng2011} corrects
the total energy by
\begin{equation} \label{eq:quad-to-lin}
    \Delta E = \sum_{n\sigma} (f_{n \sigma} - f_{n \sigma}^2) \kappa_{n \sigma},
\end{equation}
where $f_n = \ev{\rho^\sigma}{\psi_{n \sigma}}$ is the occupation of
$\ket{\psi_{n \sigma}}$ and $\kappa_{n \sigma}$ is some approximation to
$\partial^2 E / \partial f_{n \sigma}^2$. This equation assumes that
delocalization error is essentially quadratic, which was later shown to be
accurate \cite{hait2018}.
Eq.~\eqref{eq:quad-to-lin} corrects orbital energies even when
$f_{n \sigma} = 0$ or $1$ (which yields $\Delta E = 0$); while it
cannot change the total energy (and so offers no solution to DFAs'
size-inconsistency), it changes the slope of $E(N)$. Even in molecules, however,
introducing screening improves the quality of $\kappa_{n \sigma}$; later
developments in GSC \cite{zhang2015, mei2021} yielded improved atomic and
molecular photoemission spectra \cite{mei2021} in addition to orbital energies
\cite{zhang2018, yang2020b}. Localization is important in molecules because it
gives the possibility of a nonzero total energy correction. The localized orbital
scaling correction (LOSC) method initially featured localization without
screening \cite{li2018, su2020}. It improved delocalization error in the valence
orbitals of atoms and small molecules \cite{mei2021a}, with orbital energies
comparable in accuracy to molecular $GW$ quasiparticles \cite{mei2019} and
suitable as the starting point for Bethe--Salpeter equation calculations of
neutral excitations \cite{li2022d}. In addition, LOSC dramatically improves
size-consistency in the dissociation of charged diatomic molecules \cite{li2018}.

In materials, however, correcting delocalization error without localization is
impossible. Because the Bloch orbitals $\ket{\psi_{\bk n \sigma}}$ are
delocalized across all space in the thermodynamic limit, the energy is
totally insensitive to an infinitesimal change in $f_{\bk n \sigma}$ and
corrections like \eqref{eq:quad-to-lin} change neither the total energy nor
the orbital energies \cite{vlcek2015, nguyen2018}. Screening is also required
for an accurate correction. The initial LOSC method actually overcorrects
orbital energies in materials because it does not account for orbital relaxation
from other lattice electrons. \textcite{mahler2022b} adjusted LOSC by screening
the Hartree repulsion empirically; while this screened LOSC method improved band
structure prediction, its screening is identical for all systems, fundamentally
limiting its accuracy.

\begin{figure}[ht]
    \centering
    \includegraphics[width=\linewidth]{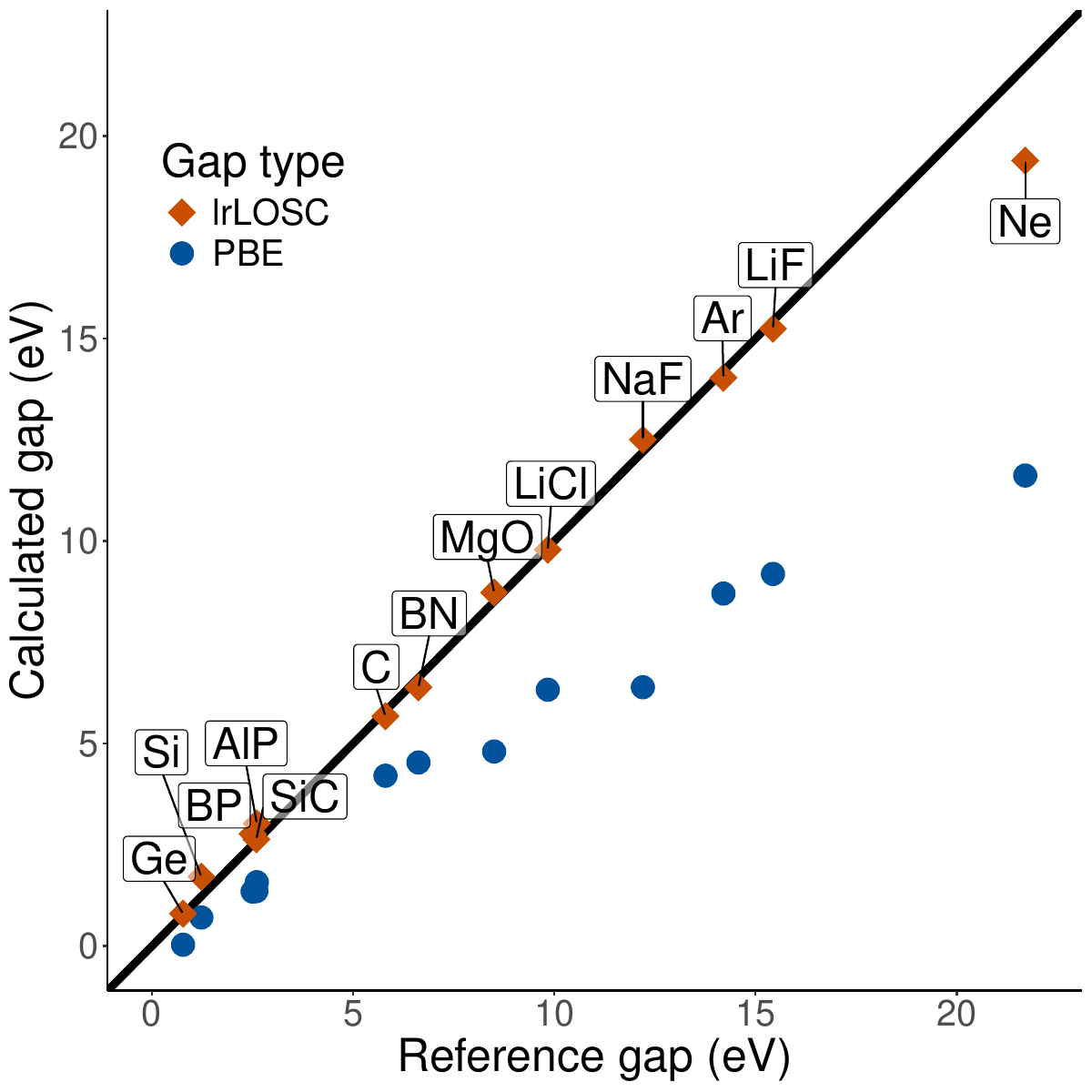}
    \caption{Fundamental gaps predicted by lrLOSC vs. by PBE. Reference:
             experimental gaps $-$ zero-point renormalization.}
    \label{fig:calcvexp}
\end{figure}

There are several related methods that leverage localization and screening to 
correct delocalization error in materials. Koopmans-compliant spectral
functionals \cite{nguyen2018, colonna2018, colonna2022}, the Wannier--Koopmans
method \cite{ma2016, weng2017, li2018b, weng2018, weng2020}, and the Wannier
optimally tuned screened range-separated hybrid functional
\cite{wing2021, ohad2022} all provide band structure corrections based on
localized orbitals---in materials, Wannier functions---and system-dependent
screening. They provide fundamental gaps and band structures with
state-of-the-art accuracy (see Table IV in the Supplemental Material \footnote{
See Supplemental Material at [URL will be inserted by publisher] for details
on the monochromatic decomposition of the curvature; a note on symmetry and
degeneracy; the unit-cell periodicity of $\kappa$; the data underlying
Fig.~\ref{fig:calcvexp}; and a comparison between lrLOSC and other methods for
correcting materials' band structures. Contains references
\cite{buberman1971, gonze1995b, palenik2016a, palenik2017a, monkhorst1976,
      rozzi2006, martyna1999, wyckoff1973, tran2009, poole1975, madelung2004,
      shishkin2007a}.
} for a comparison with the present method). However, they do not correct the
total energy of systems with a band gap \cite{nguyen2018}, which means that they
cannot restore size-consistency to DFAs.

In this work, we present lrLOSC: LOSC with linear-response screening. It
corrects semiconductors' and insulators' fundamental gaps and band structures with accuracy competitive with the previous methods. Its key difference,
however, is in its localized orbitals \cite{mahler_etal_2025}. They are constructed
from both the occupied and virtual bands, which allows lrLOSC to offer a nonzero
energy correction even for gapped systems. This total energy correction is a
necessary step toward a unified delocalization error correction applicable both
to molecules and to materials. The implementation of lrLOSC in molecules is a
parallel step towards this unification; it was found to describe both the core
and valence quasiparticle energies of small molecules accurately \cite{yu2025}.

\section{The linear-response localized orbital scaling correction}
lrLOSC modifies the DFA total energy with the ansatz
\begin{equation}
    \Delta E = \frac12 \sum_\sigma \sum_{ij \bR} \conj{\lambda}_{\bR ij\sigma}
        \left( \delta_{\bR ij} - \lambda_{\bR ij \sigma} \right)
        \kappa_{\bR ij \sigma},
\end{equation}
where $i, j, \sigma$ are respectively the band-like and spin indices of
localized orbitals $\ket{w_{\bR i \sigma}}$ and $\bR$ indexes the primitive
cell translation vectors (see below). The two ingredients of the lrLOSC energy
correction, $\lambda_{\bR ij \sigma}$ and $\kappa_{\bR ij \sigma}$, respectively
reflect the two ingredients of a delocalization error correction: localization
and screening. $\delta_{\bR ij}$ is the Kronecker delta, with
$\delta_{\bR ij} = 1$ if $i = j, \bR = \bZ$ and $\delta_{\bR ij }= 0$ otherwise;
$\conj{\lambda}$ denotes the complex conjugate of $\lambda$. Note that we
assume throughout a uniform sampling of the Brillouin zone by $N_k$ $\bk$-points
\cite{monkhorst1976}, and that the spin-polarized Kohn--Sham orbitals are
collinear:
$\braket{\psi_{\bk m\sigma}}{\psi_{\bq n\tau}} \propto
\delta_{\bk \bq} \delta_{mn} \delta_{\sigma\tau}$.

\subsection{Local occupation}
The local occupations
$\lambda_{\bR ij \sigma} =
\mel{w_{\bZ i \sigma}}{\rho^\sigma}{w_{\bR j \sigma}}$ are elements of the
density matrix
\begin{equation}
    \rho^\sigma = \sum_{\bk n} f_{\bk n \sigma} \op{\psi_{\bk n \sigma}}
\end{equation}
in the basis of localized orbitals. We note from this expression that LOSC is a
generalized Kohn--Sham density functional. LOSC's localized orbitals are called
orbitalets in finite systems \cite{su2020} and dually localized Wannier
functions (DLWFs) in periodic boundary conditions \cite{mahler_etal_2025}.
DLWFs are generalized Wannier functions \cite{wannier1937} constructed from the
Kohn--Sham Bloch orbitals $\ket{\psi_{\bk n \sigma}}$ as
\begin{equation}
    \ket{w_{\bR i \sigma}} = \frac{1}{N_k} \sum_{\bk} e^{-i \bk \cdot \bR}
        \sum_{n} U_{ni}^{\bk} \ket{\psi_{\bk n \sigma}}.
\end{equation}
The DLWFs are periodic on a supercell $N_k$ times larger than the primitive
unit cell and are indexed by a band-like index $i$ and a primitive cell
translation vector $\bR$.

At each $\bk$-point, the unitary operator $U^{\bk}$ is chosen to minimize the
dual localization cost function $F^\sigma$ \cite{gygi2003, giustino2006}, which
adds an energy localization penalty to the well-known maximally localized
Wannier function cost function \cite{marzari1997}:
\begin{equation}
    F^\sigma = 
    \sum_i 
        \left[ 
            (1 - \gamma) \Delta r_{\bZ i \sigma}^2 + 
            \gamma \Delta h_{\bZ i \sigma}^2
        \right],
\label{eq:loc}
\end{equation}
where $\Delta r_{\bZ i \sigma}^2 = \ev{r^2}{w_{\bZ i \sigma}} - \abs{\ev{\br}{w_{\bZ i \sigma}}}^2$ is the spatial variance and $\Delta h_{\bZ i \sigma}^2$
the energy variance of $\ket{w_{\bZ i \sigma}}$ \footnote{The numerical
value of $\gamma$ depends on the units chosen (see the Supplemental Material
of \cite{mahler_etal_2025} for details), but all LOSC results in both molecules and
materials so far use the same value. For $\Delta r^2$ in \si{\bohr^2} and
$\Delta h^2$ in \si{\electronvolt^2}, $\gamma = 0.47714$. Irrespective of units,
setting $\gamma = 0$ recovers maximally localized Wannier functions.}. It is
worth noting that $F^\sigma$ is a highly nonconvex functional with respect to
the localization unitaries $U^{\bk}$, often featuring local minima. We have
found that these can be avoided by sampling more $\bk$-points in the Brillouin
zone, but obtaining DLWFs is not as automatic as computing maximally localized
Wannier functions. In future work, we seek to obtain an approximation to DLWFs
that can be computed deterministically instead of iteratively minimizing a
quantity like $F^\sigma$.

The difference between DLWFs and previous implementations of $F^\sigma$ is that
they are not limited to orbitals with the same occupation; that is, they allow
the mixing of valence and conduction orbitals. Maximally localized Wannier
functions lose system-dependent information as they are constructed from more
conduction bands, while the penalty to energy variance in $F^\sigma$ produces
chemically relevant localized orbitals result even when DLWFs (or orbitalets)
are computed from both valence and high-energy conduction bands
\cite{mahler2022, yu2022a}. Mixing valence and conduction bands means that the
diagonal occupations $\lambda_{\bZ ii \sigma}$ are not constrained to integral
values, even in insulators. It is true (because the density matrix is a
projector onto the occupied manifold) that
\begin{equation}
    0 \leq \lambda_{\bZ ii \sigma} \leq 1,
\end{equation}
while general local occupations $\lambda_{\bR ij \sigma}$ can be complex, but
even a small mixing of the conduction manifold means that effectively occupied
DLWFs in gapped systems may have $\lambda_{\bZ ii}$ slightly less than $1$. It
is these noninteger local occupations that allows LOSC to yield a nonvanishing
total energy correction for nonmetals.

\subsection{Curvature: Derivation}
Expressed as an abstract operator, the lrLOSC curvature $\kappa$ for spin
$\sigma$ is
\begin{equation} \label{eq:kap-op}
\begin{split}
    \kappa^\sigma &= 
    f_{\Hxc}^{\sigma\sigma} + 
        \sum_{\nu\tau}
            f_{\Hxc}^{\sigma\nu} \chi^{\nu\tau} f_{\Hxc}^{\tau\sigma} \\ &=
    \sum_\tau \left(\epsilon^\tau\right)^{-1} f_{\Hxc}^{\tau\sigma}.
\end{split}
\end{equation}
Here, $(\epsilon^\tau)^{-1}$ is the inverse of the static microscopic
dielectric function;
\begin{equation}
\begin{split}
    f_{\Hxc}^{\sigma \nu}(\br,\br') &= 
    \frac{\delta^2 E_{\Hxc}}{\delta \rho^\sigma(\br) \delta \rho^\nu(\br')} \\
    &= \frac{1}{\abs{\br - \br'}} + 
    \frac{\delta^2 E_{\xc}}{\delta \rho^\sigma(\br) \delta \rho^\nu(\br')}
\end{split}
\end{equation}
is the Hartree--exchange-correlation kernel; and
\begin{equation}
    \chi^{\nu \tau}(\br, \br') = \frac{\delta \rho^\nu(\br)}{\delta v^\tau(\br')}
        = \frac{\delta^2 E}{\delta v^\nu(\br) \delta v^\tau(\br')}
\end{equation}
is the response of the density to an external perturbing potential $\delta v$,
to linear order. We may observe that $\kappa$ involves dielectric screening of
the bare Hartree--exchange--correlation kernel $f_{\Hxc}$; or, equivalently, it
is the sum of the bare kernel and a relaxed (screened) one, modulated by
$\chi$. 

Why do we call $\kappa$ a curvature? In the basis of canonical (Bloch) orbitals,
\begin{equation}
    \kappa_{\bk \bk' nn' \sigma} = 
    \mel{\rho_{\bk n \sigma}}
        {f_{\Hxc}^{\sigma\sigma} + 
         \sum_{\nu\tau} 
            f_{\Hxc}^{\sigma\nu} \chi^{\nu\tau} f_{\Hxc}^{\tau\sigma}
        }
        {\rho_{\bk' n' \sigma}},
\end{equation}
where $\rho_{\bk n \sigma}(\br) = \abs{\psi_{\bk n \sigma}(\br)}^2$ is the
density of the Kohn--Sham orbital $\ket{\psi_{\bk n \sigma}}$. 
\citet{yang2012a} showed that the diagonal elements of $\kappa$ in this basis
are, to second order, the curvature of the total energy $E$ with respect to the
occupation $f_{\bk n \sigma}$ of $\ket{\psi_{\bk n \sigma}}$. That is,
\begin{equation} \label{eq:kap-gsc}
\begin{split}
    \kappa_{\bk nn \sigma} &=
    \ev{f_{\Hxc}^{\sigma\sigma} +
        \sum_{\nu\tau}
            f_{\Hxc}^{\sigma\nu} \chi^{\nu\tau} f_{\Hxc}^{\tau\sigma}
       }
       {\rho_{\bk n \sigma}} \\ &=
    \frac{\partial^2 E}{\partial f_{\bk n \sigma}^2}.
\end{split}
\end{equation}
When $\ket{\psi_{\bk n \sigma}}$ is a frontier orbital,
$\partial^2 E/\partial f_{\bk n \sigma}^2$ describes the deviation of $E(N)$,
computed by a DFA, from the correct linear behavior, including the effect of
screening by all other electrons, also called orbital relaxation. This
curvature was implemented for finite systems (for which there is no $\bk$
index) in the most recent iteration of the global scaling correction
\cite{mei2021}; the same work also derived the off-diagonal extension to
\eqref{eq:kap-gsc}.

The lrLOSC ansatz is to express $\kappa$ in the DLWF basis, or more accurately
in the basis of their densities
$\rho_{\bR i \sigma}(\br) = \abs{w_{\bR i \sigma}(\br)}^2$. The matrix elements
of the lrLOSC curvature are thus
\begin{equation} \label{eq:kap-losc}
    \kappa_{\bR ij \sigma} = 
        \mel{\rho_{\bZ i \sigma}}
            {f_{\Hxc}^{\sigma \sigma} + \sum_{\nu \tau} 
                f_{\Hxc}^{\sigma \nu} \chi^{\nu \tau} f_{\Hxc}^{\tau \sigma}}
            {\rho_{\bR j \sigma}}.
\end{equation}
(Due to translational symmetry, only one $\bR$ index, an offset, is required to
keep track of the DLWF unit cells; without loss of generality, the left DLWF
density always has $\bR = \bZ$.) Even if only the diagonal elements of $\kappa$
are considered in the basis of canonical orbitals, the unitary transformation to
the DLWF basis will produce off-diagonal elements for which $i \neq j$ or
$\bR \neq \bZ$. These off-diagonal curvature elements allow interactions
between pairs of orbitals to affect the total energy, even at long range. They
were first used to ensure a correction to the total energy necessary to describe
the dissociation energy curves of finite systems correctly \cite{li2018}; they
also improve the description of molecules' core-level binding energies
\cite{yu2025} and, in lrLOSC, offer a supporting example of the need for
localized orbitals in any delocalization error correction for extended systems
\cite{nguyen2018}. Note that analogous elements for canonical orbitals were
derived as cross-terms $\partial^2 E / \partial f_{n\sigma} \partial f_{m\sigma}$
in the Supporting Information of \cite{mei2021}. 

\subsection{Curvature: implementation}
Computing $\kappa_{\bR ij \sigma}$ na\"ively looks computationally difficult
because of the linear-response function $\chi^{\nu\tau}$, which is nonlocal in
both real and reciprocal space. Several computational discoveries allow us to
obtain it in a reasonable amount of time. We can leverage density functional
perturbation theory \cite{baroni2001} to treat linear-response quantities in
terms of monochromatic components indexed by a wavevector $\bq$ (in our work,
$\{\bq\} = \{\bk\}$). This decreases the computational cost of computing
$\kappa$ by a factor of $N_k$. In particular, \textcite{timrov2018} showed
that linear and linear-response operators decompose monochromatically, and
\textcite{colonna2022} demonstrated that Wannier function densities do the
same. We can therefore write $\kappa_{\bR ij \sigma}$, utilizing the notation
from \cite{colonna2022}, as
\begin{widetext}
\begin{equation} \label{eq:kap-lr}
\begin{split}
    \kappa_{\bR ij \sigma} &=
    \frac{1}{N_k} \sum_{\bq} 
        e^{-i\bq\cdot\bR} \kappa_{\bZ ij \sigma}^{\bq} \\ &=
    \frac{1}{N_k} \sum_{\bq} e^{-i\bq\cdot\bR}
        \left[ 
            \mel{\rho_{\bZ i \sigma}^{\bq}}
                {f_{\Hxc}^{\bq; \sigma\sigma} + \sum_{\nu\tau} 
                    f_{\Hxc}^{\bq; \sigma\nu} 
                    \chi^{\bq; \nu\tau} 
                    f_{\Hxc}^{\bq; \tau\sigma}
                }
                {\rho_{\bZ j \sigma}^{\bq}}
        \right] \\ &=
    \frac{1}{N_k} \sum_{\bq} e^{-i\bq\cdot\bR} 
        \left[
            \braket{\rho_{\bZ i \sigma}^{\bq}}{V_{\bZ j \sigma}^{\bq}} +
            \sum_\tau
                \braket{\delta\rho_{\bZ i \tau}^{\bq}}{V_{\bZ j \tau}^{\bq}}
        \right],
\end{split}
\end{equation}
\end{widetext}
where
\begin{equation}
    V_{\bZ j \tau}^{\bq}(\br) = 
    \int d\br'\,
        f_{\Hxc}^{\bq; \tau \sigma}(\br, \br') \rho_{\bZ j \sigma}^{\bq}(\br')
\end{equation}
and
\begin{multline}
    \delta \rho_{\bZ i \tau}^{\bq}(\br) =
        \sum_\nu \int d\br'\, 
            \chi^{\tau \nu}(\br,\br') V_{\bZ i \nu}^{\bq}(\br') \\ =
    \sum_\nu \iint d\br'\, d\br''\, 
        \chi^{\tau\nu}(\br, \br')
        f_{\Hxc}^{\bq; \nu\sigma}(\br', \br'')
        \rho_{\bZ j \sigma}^{\bq}(\br'').
\end{multline}
The Supplemental Material provides more detail on this derivation. We note that
the diagonal elements $\kappa_{\bZ ii \sigma}^{\bq}$ are identical to the
relaxed (numerator) terms in the screening coefficients of the
Koopmans-compliant Wannier method \cite{colonna2022}.

This form makes clear that $\kappa_{\bZ ij \sigma}^{\bq}$ is the sum of a
bare curvature $\braket{\rho_{\bZ i \sigma}^{\bq}}{V_{\bZ j \sigma}^{\bq}}$ and
a screened term $\braket{\delta\rho_{\bZ i \tau}^{\bq}}{V_{\bZ j \tau}^{\bq}}$.
The density-density linear response function $\chi^{\tau\nu}$ does not appear
explicitly, but its effects are contained in
$\ket{\delta\rho_{\bZ i \tau}^{\bq}}$. We compute
$\ket{\delta\rho_{\bZ i \tau}^{\bq}}$ using the iterative Sternheimer equation.
True to their role in moderating the LOSC correction, the screened terms are
typically of opposite sign and smaller in magnitude to the bare terms.

Computing $\ket{\delta \rho_{\bZ i \tau}^{\bq}}$ occupies the bulk of lrLOSC's
computational time. For each $\ket{\rho_{\bZ i \tau}^{\bq}}$, there are
$N_{\occ}$ coupled equations that must be solved at each $\bk$-point, for a
total runtime scaling as $\Ord(N_{\text{DLWF}} N_{\bk}^2 N_\occ^3)$ (since
$N_{\bq} = N_{\bk}$).

\subsection{Hamiltonian correction}
The LOSC correction to the Hamiltonian is derived by the chain rule and the
generalized complex (Wirtinger) derivative \cite{wirtinger1927}, since
\begin{equation}
\begin{split}
    \Delta h^\sigma &= 
    \frac{\delta \Delta E^\sigma}{\delta \rho^\sigma} \\ &=
    \sum_{\bR ij} 
        \left[ 
            \frac{\partial \Delta E^\sigma}{\partial \lambda_{\bR ij \sigma}}
            \frac{\delta \lambda_{\bR ij \sigma}}{\delta \rho^\sigma} +
            \frac{\partial \Delta E^\sigma}
                 {\partial \conj{\lambda}_{\bR ij \sigma}}
            \frac{\delta \conj{\lambda}_{\bR ij \sigma}}{\delta \rho^\sigma}
        \right];
\end{split}
\end{equation}
it can be shown that
\begin{equation}
    \mel{w_{\bZ i \sigma}}{\Delta h^\sigma}{w_{\bR j \sigma}} =
    \left(
        \frac12 \delta_{\bR ij} - \lambda_{\bR ij \sigma}
    \right)
        \Re{\kappa_{\bR ij \sigma}}.
\end{equation}

We compute corrections to the Bloch orbital energies by transforming $\Delta h^\sigma$ to the Bloch basis and diagonalizing $h + \Delta h^\sigma$. It should
be noted that this produces new Bloch functions and orbital energies that are no
longer strictly Kohn--Sham eigenfunctions; this is observed to break some
band degeneracies, as seen below. The orbital energy correction is currently
implemented as a one-shot procedure. In future work, we will update the Bloch
functions and their energy eigenvalues self-consistently, which should
ameliorate the degeneracy breaking; this has already been implemented for
molecules \cite{mei2020a}.

\section{Results}
We compute the fundamental gaps and band structures of thirteen semiconductors
and insulators. Our DFA calculations are performed with
\texttt{Quantum ESPRESSO} \cite{giannozzi2009, giannozzi2017, giannozzi2020},
version 7.1. We use the PBE functional \cite{perdew1996}, optimized
norm-conserving Vanderbilt pseudopotentials \cite{hamann2013} built under PBE
with scalar relativistic corrections, a wavefunction kinetic energy cutoff of
\SI{75}{\rydberg}, and a $6 \times 6 \times 6$ Monkhorst--Pack sampling of the
irreducible Brillouin zone; the PBE valence band maximum and conduction band
minimum is converged to within \SI{0.01}{\electronvolt}. We scale the
correction \cite{gygi1986} to the Coulomb divergence in
$\braket{\delta\rho_{\bZ i \tau}^{\bq}}{V_{\bZ j \tau}^{\bq}}$ by the
macroscopic dielectric constant $\epsilon_\infty$; it is computed in
\texttt{Quantum ESPRESSO}'s \texttt{PHonon} module, with the same parameters,
which are sufficient to converge $\epsilon_\infty$ to 0.01.

The DLWFs are constructed from all valence bands provided by the
pseudopotentials, in addition to twice as many conduction bands as the
connectivity of the lattice. For example, an atom in a diamond lattice (\ce{C},
\ce{Si}, \ce{SiC}, \ce{Ge}) has four nearest neighbors, so we use eight
conduction bands in the construction of the DLWFs. These are disentangled with
the procedure of \textcite{souza2001} to a composite set containing half as
many conduction bands (equal to lattice connectivity, e.g. $4$ for the diamond
lattice). A locally maintained fork \cite{mahler_etal_2025} of \texttt{wannier90}
\cite{mostofi2008, mostofi2014, pizzi2020} is used to calculate the Wannier
functions.

lrLOSC is implemented in a local fork of \texttt{Quantum ESPRESSO}, version 7.2
\footnote{\url{https://gitlab.oit.duke.edu/jzw5/qe-lrlosc}}. The monochromatic
decomposition of $\ket{\rho_{\bZ i \sigma}^{\bq}}$, as well as the flow of the LOSC
calculation (localization, followed by curvature, followed by corrections to the
spectrum and the total energy), are adapted from the \texttt{KCW} module
\cite{colonna2022}. We leverage existing linear-response routines in
\texttt{Quantum ESPRESSO} to compute $\ket{\delta\rho_{\bZ i \tau}^{\bq}}$ via
the Sternheimer equation.
The data that support our findings are openly available 
at the Duke Research Data Repository \cite{williams2025}.

There is only one free parameter, $\gamma$, in lrLOSC; it describes the balance
between spatial and energy localization of the DLWFs in \eqref{eq:loc}. The
value we choose, $\gamma = 0.47714$, was originally optimized for LOSC in finite
systems \cite{su2020}; it is also used for sLOSC in materials
\cite{mahler2022b}. This choice of $\gamma$ has also been shown for molecules to
provide accurate corrections for several starting DFAs \cite{su2020, yu2025}.
We have thus used it for lrLOSC without further modification.

lrLOSC greatly improves fundamental gaps relative to PBE. All the systems
compared, except neon and argon, have zero-point renormalization (ZPR)
corrections available in the literature
\cite{miglio2020, shang2021, engel2022}, which allows us to compare the
\emph{electronic} (fundamental) contribution to the experimental (optical) band
gap to lrLOSC's prediction. We omit Ne and Ar from the quantitative discussion
of lrLOSC. Calculated from the remaining eleven materials, the mean absolute
error (MAE) in the lrLOSC fundamental gap relative to the electronic
(experimental $-$ ZPR) gap, is \SI{0.22}{\electronvolt}; compare this to PBE's
MAE of \SI{2.14}{\electronvolt}. This result is comparable to results reported
by other delocalization error correction methods, and even outperforms some
many-body methods.

\begin{table}[htb]
\caption{Reported mean absolute errors (MAE) of the fundamental gap. For a
comparison limited to the same materials in every method, see the Supplemental
Material.)}
\label{tab:mae}
\begin{ruledtabular}
\begin{tabular}{ccc}
Method  & Reference & MAE (\si{\electronvolt}) \\
\colrule
PBE         & This work         & 2.14 \\
lrLOSC      & This work         & 0.22 \\
$G_0W_0$    & \cite{chen2015a}  & 0.45 \\
qs$GW$      & \cite{chen2015a}  & 0.62 \\
EOM-CCSD    & \cite{vo2024}     & 0.42 \\
Koopmans-compliant integral         & \cite{nguyen2018} & 0.27 \\
Wannier--Koopmans        & \cite{weng2017}   & 0.45 \\
WOT-SRSH    & \cite{wing2021}   & 0.08 \\
\end{tabular}
\end{ruledtabular}    
\end{table}

\begin{figure*}[ht!]
    \centering
    \includegraphics[width=\textwidth]{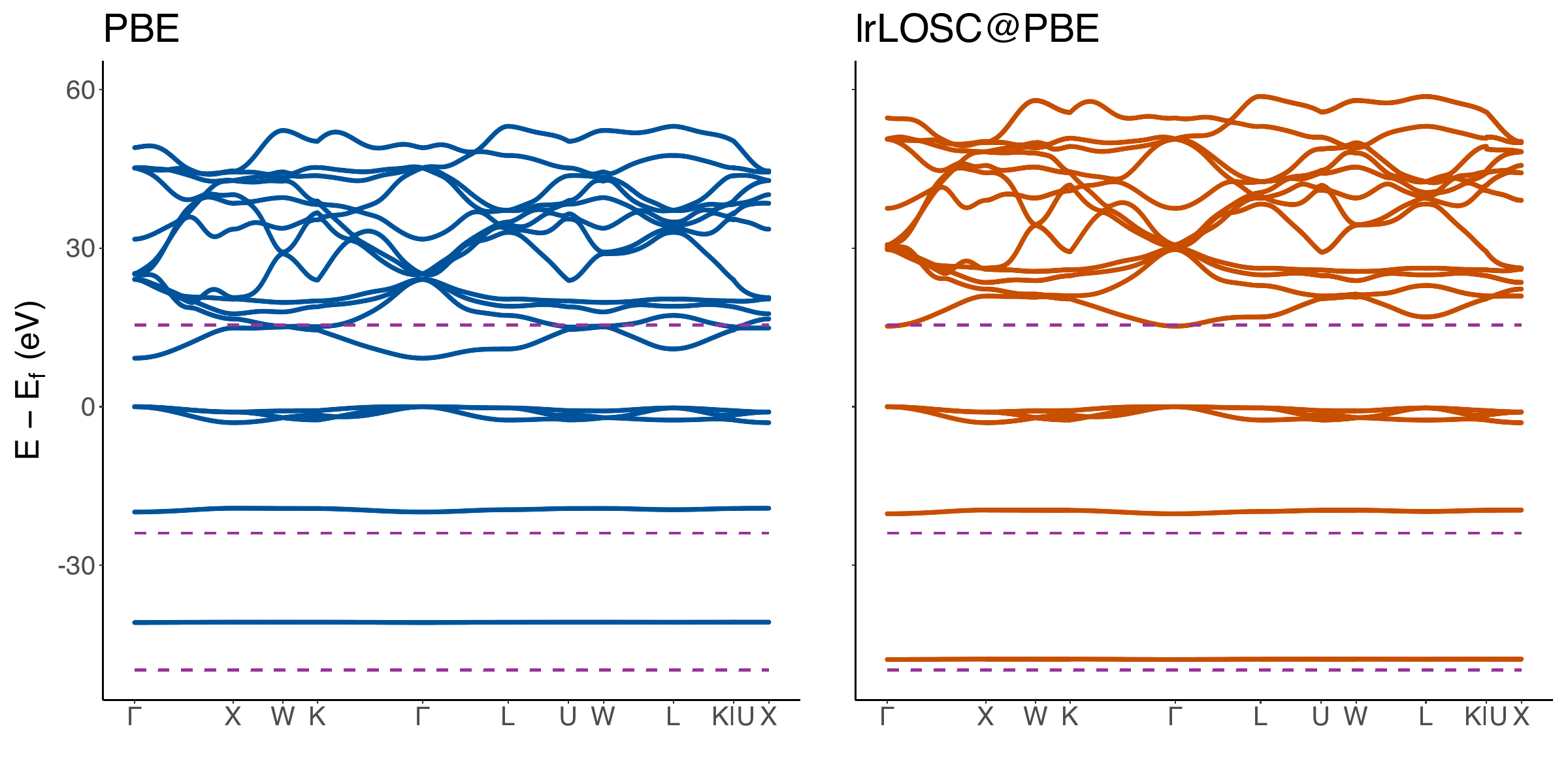}
    \caption{\ce{LiF} band structure by PBE and lrLOSC. Purple dashed lines (top to
             bottom): Experimental gaps adjusted by ZPR; experimental energies
             for F \ce{2s}, Li \ce{1s} states.}
    \label{fig:lif}
\end{figure*}

In Fig.~\ref{fig:lif}, we show the band structure of lithium fluoride calculated
with PBE (left) and lrLOSC (right). The DFA gap is only \SI{9.19}{\electronvolt},
much smaller than the electronic gap (\SI{15.43}{\electronvolt}, including a
phonon-mediated zero-point renormalization of \SI{-1.231}{\electronvolt}).
lrLOSC shifts the occupied bands down and the virtual bands up relative to PBE
and predicts a gap of \SI{15.24}{\electronvolt}, within \SI{0.2}{\electronvolt}
(and $1\%$) of the correct gap. lrLOSC also improves the core-level energies,
especially the lithium \ce{1s} state, compared to the parent DFA. Its performance is
very similar in this system to the Koopmans-compliant Wannier functional
\cite{colonna2022}; it slightly outperforms $G_0W_0$ in the electronic gap, but
is outperformed in turn for the core-level energies. It is worth noting that all
the computational methods still yield a higher energy for the core electrons than
is experimentally observed. We hypothesize that this is a limitation of
describing (semi)core orbitals with pseudopotentials: electrons closest to the
nucleus will be most sensitive to the nonsmoothness of the potential there.

\begin{table}[hb]
\centering
\caption{\label{tab:lif}%
         Theoretical and experimental core-level quasiparticle energies and
         electronic gap of LiF (all in eV). $G_0W_0$ is from \cite{wang2003},
         and uses LDA rather than PBE; Koopmans spectral functional (KI) is from
         \cite{colonna2022}; PBE and lrLOSC are from this work. Experimental
         core energies are from \cite{johansson1976}; the gap is from
         \cite{heyd2005} and containing references, and the zero-point
         renormalization is computed in \cite{engel2022}.}
\begin{ruledtabular}
\begin{tabular}{rrrrrr}
     & PBE & $G_0 W_0$ & KI & lrLOSC & Exp. \\ 
    \colrule
    \ce{Li} (\ce{1s}) & $-40.8$ & $-47.2$ & $-46.6$ & $-47.8$ & $-49.8$ \\
    \ce{F} (\ce{2s}) & $-19.5$ & $-24.8$ & $-19.5$ & $-19.7$ & $-23.9$ \\
    $E_g-$ ZPR&   $9.2$ &  $14.3$  & $15.3$ &  $15.2$ &  $15.4$ \\
\end{tabular}
\end{ruledtabular}
\end{table}

\begin{figure*}[htb]
    \centering
    \includegraphics[width=\textwidth]{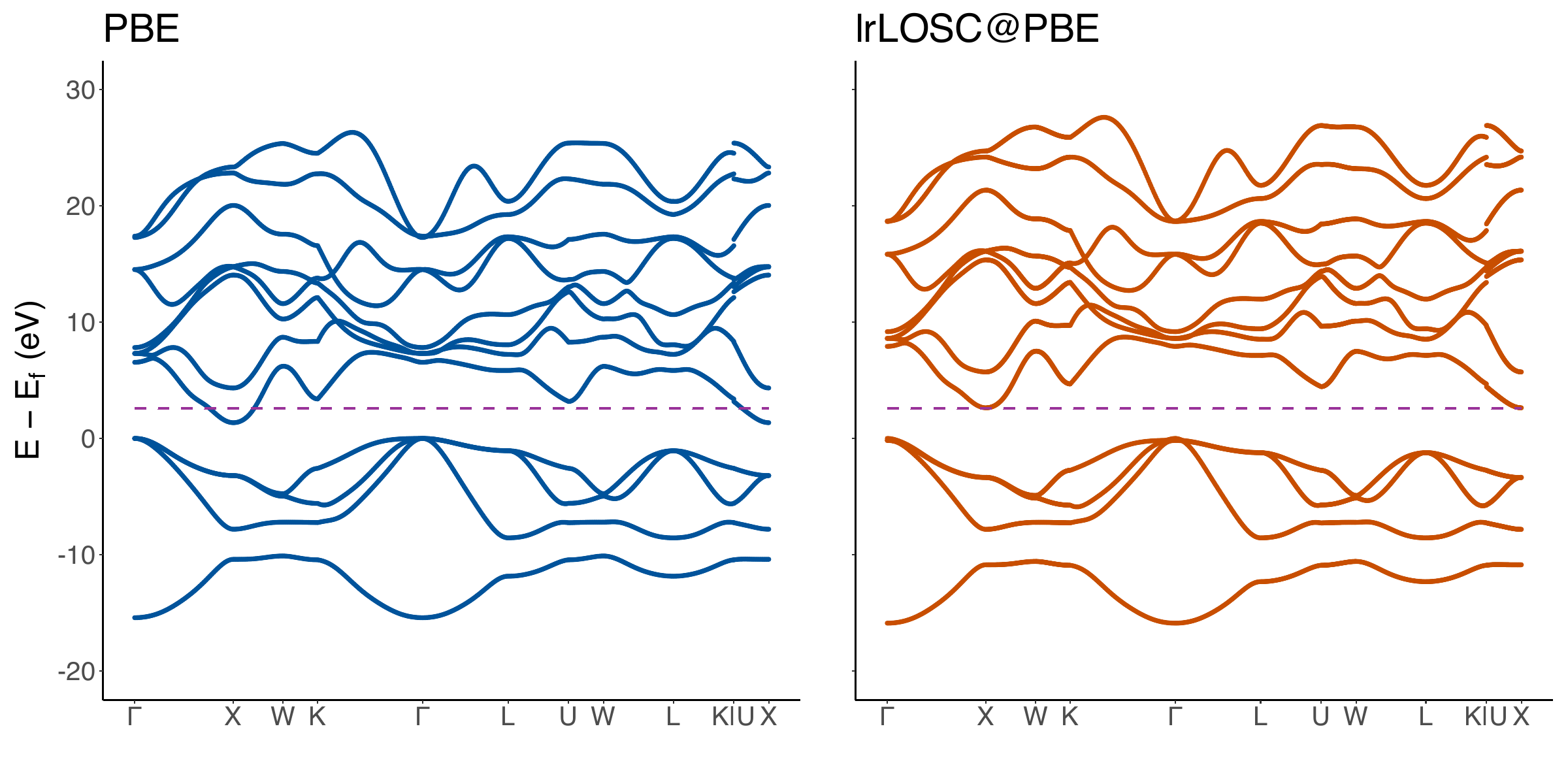}
    \caption{SiC band structure, with the Fermi energy set to zero. Purple
             dashed line: Experimental gap adjusted for ZPR.}
    \label{fig:sic}
\end{figure*}

The lrLOSC band gap correction for silicon carbide is particularly accurate
(Fig.~\ref{fig:sic}). The qualitative features of the DFA band gap are virtually
unchanged, but the electronic gap is adjusted to within \SI{0.04}{\electronvolt}
of the correct value. A modest increase in conduction band energies---coupled
with a large downward shift in core-level energies, leaving valence energies
mostly unchanged relative to the Fermi level---is typical behavior for lrLOSC on
the systems tested. In the systems tested, however, the lrLOSC Fermi level is
always observed to be lower than that of PBE. We note also that lrLOSC
breaks the degeneracy between the highest- and second-highest-energy
occupied bands at the $\Gamma$ point by about \SI{0.2}{\electronvolt}
(Table \ref{tab:sic}). A similar splitting of degenerate bands is observed for
the semicore orbitals in NaF and Si (see the ``Note on symmetry and degeneracy"
in the Supplemental Material). This is likely because we do not use degenerate
perturbation theory when computing the curvature; however, its effect is small
in the systems we have studied.

\begin{table}[htb]
\centering
\caption{\ce{SiC} orbital energies of occupied bands at $\Gamma$.}
\label{tab:sic}
\begin{ruledtabular}
\begin{tabular}{@{}rrr@{}}
    Band & PBE & lrLOSC \\ 
    \colrule
    VBM$-3$ & $-5.9099$ & $-7.1407$ \\
    VBM$-2$ &  $9.5027$ &  $8.5658$ \\
    VBM$-1$ &  $9.5027$ &  $8.5821$ \\
    VBM     &  $9.5027$ &  $8.7415$ \\
\end{tabular}
\end{ruledtabular}
\end{table}

\section{Conclusion}
Because it includes both localization and system-dependent screening, lrLOSC
corrects delocalization error effectively in both molecules and materials. It
provides accurate, size-consistent orbital and band energy corrections in
semiconductors and insulators, matching the performance of many-body
perturbation theory without requiring any many-body observables. In its current
implementation, it is limited to gapped systems, although its extension to
metals should parallel that for other density functional perturbation theory
methods \cite{baroni2001}. Future work will implement an approximate, but still
system-dependent, curvature, improving LOSC's computational efficiency. We also
seek to apply LOSC self-consistently, yielding a correction to the electron
density; this feature has already been demonstrated for molecules
\cite{mei2020a}. With these updates, LOSC will be applicable to interfaces,
including molecules on a solid surface. Modeling energy level alignment at
interfaces is a major challenge for electronic structure methods
\cite{flores2009, egger2015, liu2020d}, and lrLOSC promises a nearly
parameter-free solution within the confines of generalized Kohn--Sham DFT.

\begin{acknowledgments}
We thank Yichen Fan and Jincheng Yu
for helpful discussions of LOSC in molecules.
We gratefully acknowledge funding
from the National Science Foundation (CHE-2154831)
and the National Institutes of Health (5R01GM061870-20).
\end{acknowledgments}

\bibliography{lrLOSC}

\end{document}


\title{Supplemental Material:\ Localized Orbital Scaling Correction with
       Linear Response in Materials}
\author{Jacob Z. Williams}
\altaffiliation{Present address:~Theoretical Division, 
                Los Alamos National Laboratory, Los Alamos, NM 87545, USA}
\affiliation{Department of Chemistry, Duke University, Durham, NC 27708, USA}
\author{Weitao Yang}
\email{weitao.yang@duke.edu}
\affiliation{Department of Chemistry, Duke University, Durham, NC 27708, USA}
\affiliation{Department of Physics, Duke University, Durham, NC 27708, USA}
\date{\today}

\maketitle

\section{Monochromatic implementation of the curvature}
The second derivative of the Kohn--Sham total energy with respect to the
occupation of the canonical orbitals has matrix elements
\cite{yang2012, mei2021}
\begin{equation}
    \frac{\partial^2 E}{\partial f_{n\sigma}^2} 
        = \kappa_{nn}^\sigma 
        = \mel{\rho_{n\sigma}}
              {f_{\Hxc}^{\sigma \sigma} + \sum_{\nu \tau} 
                f_{\Hxc}^{\sigma \nu} \chi^{\nu \tau} f_{\Hxc}^{\tau \sigma}}
              {\rho_{n\sigma}}.
\end{equation}
Here, $f_{n\sigma}$ is the occupation of the canonical orbital
$\ket{\psi_{n\sigma}}$; $\rho_{n\sigma}(\br) = \abs{\psi_{n\sigma}(\br)}^2$ is
its density;
\begin{equation}
    f_{\Hxc}^{\sigma \tau}(\br,\br') 
        = \frac{\delta^2 E_{\Hxc}}{\delta\rho^\sigma(\br)\delta\rho^\tau(\br')}
        = \frac{1}{\abs{\br-\br'}} +
          \frac{\delta^2 E_{\xc}}{\delta\rho^\sigma(\br)\delta\rho^\tau(\br')}
\end{equation}
is the Hartree--exchange-correlation kernel; and
\begin{equation}
    \chi^{\nu \tau}(\br, \br') = \frac{\delta \rho^\nu(\br)}{\delta v^\tau(\br')} 
        = \frac{\delta^2 E}{\delta v^\nu(\br) \delta v^\tau(\br')}
\end{equation}
is the density-density response function, which encodes screening. We assume
that the Kohn--Sham functional is semilocal for compactness of notation.

In LOSC, we express $\kappa_\sigma$ in the the dually localized Wannier function (DLWF) basis \cite{mahler_etal_2025}
to apply it to LOSC. In this basis, its matrix elements become
\begin{equation}
    \kappa_{\bR ij}^\sigma = \mel{\rho_{\bZ i \sigma}}{f_{\Hxc}^{\sigma \sigma} +
        \sum_{\nu \tau} f_{\Hxc}^{\sigma \nu} \chi^{\nu \tau} f_{\Hxc}^{\tau \sigma}}
        {\rho_{\bR j \sigma}}.
\end{equation}

To simplify this, define $\ket{V_{\bR j \tau}}$ and
$\ket{\delta \rho_{\bZ i \tau}}$ by
\begin{equation}
\begin{split}
    V_{\bR j \tau}(\br) &= \int d\br'\, f_{\Hxc}^{\tau \sigma}(\br, \br') 
        \rho_{\bR j \sigma}(\br'); \\
    \delta \rho_{\bZ i \tau}(\br) &= \sum_\nu \int d\br'\, \chi^{\tau \nu}(\br, \br') 
        V_{\bZ i}^\nu(\br') 
        = \sum_\nu \int d\br'\, \chi^{\tau \nu}(\br, \br') \int d\br''\, 
            f_{\Hxc}^{\nu \sigma}(\br', \br'') \rho_{\bZ i \sigma}(\br'');
\end{split}
\end{equation}
then
\begin{equation}
    \kappa_{\bR ij}^\sigma
        = \braket{\rho_{\bZ i \sigma}}{V_{\bR j \sigma}} + \sum_\tau
          \braket{\delta \rho_{\bZ i \tau}}{V_{\bR j \tau}}.
\end{equation}
(Note that $\sigma$ is a fixed spin index, while $\tau$ and $\nu$ vary.)

We call $\ket{V_{\bR j \sigma}}$ the bare potential from DLWF
$\ket{w_{\bR j \sigma}}$ and $\ket{\delta \rho_{\bZ i \tau}}$ the 
screened response of $\ket{\rho_{\bZ i \sigma}}$ to $\ket{V_{\bZ i \tau}}$.

\subsection{The monochromatic decomposition}
Linear-response quantities in periodic boundary conditions are expressible as a
sum over crystal momentum vectors $\bq$. If the Brillouin zone is uniformly
sampled by a Monkhorst--Pack mesh \cite{monkhorst1976}, the $\bq$-points
decouple from one another; in the language of density functional perturbation
theory, only \emph{monochromatic} (i.e., single-$\bq$) perturbations need be
considered \cite{baroni2001}. \citet{timrov2018} leveraged this feature in the
context of DFT+$U$, monochromatically decomposing the linear-response
formulation of the Hubbard parameter. More recently, \citet{colonna2022} found
that Wannier function densities---including DLWF densities
$\ket{\rho_{\bR i \sigma}}$---also decompose monochromatically. At the risk of
abuse of notation, we alternate freely between states like
$\ket{\rho_{\bR i \sigma}}$ and their real-space representations
$\rho_{\bR i \sigma}(\br) = \braket{\br}{\rho_{\bR i \sigma}}$.

Assuming identical Monkhorst--Pack meshes $\{\bk\}$ and $\{\bq\}$, both
centered at the origin $\Gamma$ of the Brillouin zone, and using $\conj{z}$ for
the complex conjugate of $z$, \citet{colonna2022} found that
\begin{equation}  \label{eq:mono-rho}
    \rho_{\bR i \sigma}(\br) =
    \abs{w_{\bR i \sigma}(\br)}^2 =
    \frac{1}{N_k} \sum_{\bq} e^{i\bq\cdot\br} e^{i\bq\cdot\br} 
        \rho_{\bR i \sigma}^{\bq}(\br) =
    \frac{1}{N_k} \sum_{\bq} e^{i\bq\cdot\br} e^{-i\bq\cdot\bR}
        \rho_{\bZ i \sigma}^{\bq}(\br),
\end{equation}
where 
\begin{equation}
    \rho_{\bR i \sigma}^{\bq}(\br) =
    e^{-i \bq \cdot \bR} \times \frac{1}{N_k} \sum_{\bk}
      \conj{\varphi}_{\bk i \sigma}(\br) \varphi_{(\bk + \bq) i \sigma}(\br).
\end{equation}
$\ket{\varphi_{\bk i \sigma}} = \sum_n U_{ni}^{\bk} \ket{u_{\bk n \sigma}}$ is
the periodic part of the Kohn--Sham (Bloch) orbital transformed by the
unitary localization matrix (but not the Fourier transform into the Wannier
basis); the Kohn--Sham orbitals themselves are given by
$\psi_{\bk n \sigma}(\br) = e^{i\bk\cdot\br} u_{\bk n \sigma}(\br)$. The
quantities $\ket{\rho_{\bZ i \sigma}^{\bq}}$ are periodic on the primitive cell.
Additionally, note from Eq.~\eqref{eq:mono-rho} that the unit cell offset
$\bR$ appears only as a phase; only $\ket{\rho_{\bZ i}^{\bq}}$ need ever be
computed explicitly.

The monochromatic decomposition extends to $\ket{V_{\bR j \tau}}$ and
$\ket{\delta \rho_{\bZ i \tau}}$, since
\begin{equation}
\begin{split}
    V_{\bR j \tau}(\br) 
        &= \int d\br'\, f_{\Hxc}^{\tau \sigma}(\br, \br') 
            \rho_{\bR j \sigma}(\br') \\
        &= \frac{1}{N_k} \sum_{\bq} e^{i\bq \cdot \br} e^{-i \bq \cdot \bR}
            \int d\br'\, f_{\Hxc}^{\tau \sigma}(\br, \br')
            \rho_{\bZ j \sigma}^{\bq}(\br') \\
        &= \frac{1}{N_k} \sum_{\bq} e^{i\bq \cdot \br} e^{-i \bq \cdot \bR} \,
            V_{\bZ j \tau}^{\bq}(\br)
\end{split}
\end{equation}
and
\begin{equation}
\begin{split}
    \delta \rho_{\bZ i \tau}(\br)
        &= \sum_\nu \int d\br'\, \chi^{\tau \nu}(\br, \br') V_{\bZ j \nu}(\br') \\
        &= \frac{1}{N_k} \sum_\nu \int d\br'\, \chi^{\tau \nu}(\br, \br') 
            \sum_{\bq} e^{i\bq \cdot \br'} V_{\bZ i \nu}^{\bq}(\br') \\
        &= \frac{1}{N_k} \sum_{\bq} e^{i \bq \cdot \br} \sum_\nu \int d\br'\,
            e^{-i \bq \cdot (\br' - \br)} \chi^{\tau \nu}(\br, \br')
            V_{\bZ i \nu}^{\bq}(\br') \\
        &= \frac{1}{N_k} \sum_{\bq} e^{i\bq \cdot \br} \sum_\nu \int d\br'\, 
            \chi^{\tau \nu; \bq}(\br, \br) V_{\bZ i \nu}^{\bq}(\br') \\
        &= \frac{1}{N_k} \sum_{\bq} e^{i\bq \cdot \br}
            \delta \rho_{\bZ i \tau}^{\bq}(\br).
\end{split}
\end{equation}
When decomposing $\ket{V_{\bR j \tau}}$, we used the fact that
$f_{\Hxc}^{\tau \sigma}$ is periodic on the primitive cell, and for
$\ket{\delta \rho_{\bZ i \tau}}$ we used that
\begin{equation}
    \chi^{\tau \nu}(\br, \br') = \sum_{\bq} e^{i \bq \cdot (\br' - \br)}
        \chi^{\tau \nu; \bq}(\br, \br').
\end{equation}

Thus, we obtain at last that
\begin{equation}
    \kappa_{\bR ij}^\sigma 
    = \frac{1}{N_k} \sum_{\bq} e^{- i \bq \cdot \bR} \left[
        \braket{\rho_{\bZ i \sigma}^{\bq}}{V_{\bZ j \sigma}^{\bq}} + \sum_\tau
        \braket{\delta \rho_{\bZ i \tau}^{\bq}}{V_{\bZ j \tau}^{\bq}}
      \right].
\end{equation} 

In practice, $\ket{\delta \rho_{\bZ i \tau}^{\bq}}$ is computed iteratively,
instead of evaluating $\chi^{\tau \nu}$ directly. Extending a standard result
in perturbation theory \cite{baroni2001}, we have that
\begin{equation}
    \delta \rho_{\bZ i \tau}^{\bq}(\br) = \sum_{\bk} \sum_{n \in \occ}
        \conj{\psi}_{\bk n \tau}(\br) \delta\psi_{(\bk + \bq)n \tau}(\br)
        + \text{c.c.},
\label{eq:drho}
\end{equation}
where
\begin{equation}
    \delta \psi_{\bk n \tau}(\br) = \sum_{m \neq n} \psi_{\bk m \tau}(\br)
        \frac{\mel{\psi_{\bk m \tau}}{\delta V^\tau}{\psi_{\bk n \tau}}}
             {\eps_{\bk n \tau} - \eps_{\bk m \tau}}
\label{eq:dpsi}
\end{equation}
is the linear variation in the Kohn--Sham orbital $\ket{\psi_{\bk n \tau}}$
under the perturbing potential $\delta V^\tau$. Computing
$\ket{\delta \psi_{\bk n \tau}}$ by Table~\ref{eq:dpsi} requires a double sum
over both occupied and unoccupied orbitals, which is very expensive; however,
it is known \cite{baroni2001} that perturbations including only occupied
orbitals do not contribute to the density variation. The iterative Sternheimer
equations, one for each $\bk$-point and spin index, yield the variations
$\ket{\delta \psi_{(\bk + \bq) n \tau}}$ in the occupied orbitals as solutions to
\begin{equation}
    \left[
        h^{\tau} + \gamma P_{\occ}^{(\bk + \bq) \tau} - \eps_{\bk n \tau}
    \right] \delta \psi_{(\bk + \bq) n \tau}(\br) \\
    = - P_{\vir}^{(\bk + \bq) \tau}
    \left[
        V_{\bZ i \tau}^{\bq}(\br) + \delta V_{\bZ i \tau}^{\bq}(\br)
    \right] \psi_{\bk n \tau}(\br).
\label{eq:stern}
\end{equation}
In \eqref{eq:stern}, $P_{\occ}^{\bk \tau}$ is the projector onto the occupied
Kohn--Sham states with spin $\tau$ at $\bk$; likewise,
$P_{\vir}^{\bk \tau} = I - P_{\occ}^{\bk}$ is the projector onto the virtual
states (it's used as an orthogonalizer). The small, positive real number $\gamma$
serves as a preconditioner, and the potential response
\begin{equation}
    \delta V_{\bZ i \tau}^{\bq}(\br) 
        = \sum_{\varsigma} \int d\br'\, f_{\Hxc}^{\tau \varsigma}(\br, \br') 
            \delta \rho_{\bZ i \varsigma}^{\bq}(\br')
\label{eq:dv}
\end{equation}
is updated at each iteration. Eqs.\ \eqref{eq:drho}, \eqref{eq:stern}, and
\eqref{eq:dv} provide a self-consistent cycle for obtaining
$\ket{\delta \rho_{\bZ i \tau}^{\bq}}$, from which the LOSC curvature is
obtained.

The purpose of the monochromatic decomposition is computational savings.
Computing $\ket{\delta \rho_{\bZ i}}$ na\"ively scales as $(N_e N_k)^3$, while
calculating each $\ket{\delta \rho_{\bZ i \sigma}^{\bq}}$ scale as $N_k N_e^3$;
since there are $N_k$ of them, a factor of $N_k$ is saved overall. As the
linear-response computation is still quite expensive, this is a significant time
savings. For the systems tested in this work, $N_k = 216$, so computing lrLOSC
on the large supercell would require roughly two orders of magnitude more
computational effort.

\subsection{Monochromatic implementation of sLOSC curvature}
LOSC initially approximated the curvature as
\begin{equation}
\begin{split}
    \kappa_{\bR ij}^{\sigma}(\beta)
        &= J_{\bR ij} + \beta X_{\bR ij}^\sigma \\
        &= \iint d\br\, d\br'\, 
             \frac{\rho_{\bZ i \sigma}(\br) \rho_{\bR j \sigma}(\br')}
                  {\abs{\br - \br'}}
           + \beta \int d\br\, C_X \left[ \rho_{\bZ i \sigma}(\br) 
             \rho_{\bR j \sigma}(\br) \right]^{2/3},
\end{split}
\end{equation}
where $\beta = 6(1 - 1/\sqrt[3]{2})$ is chosen so that the self-exchange in any
one-electron density is twice that of the same density with half an electron.
$C_X = 0.75\sqrt[3]{6/\pi}$ is the Dirac exchange constant \cite{li2018,su2020}.

This curvature corrects delocalization error in small molecules effectively, but
the screening of the Coulomb repulsion by lattice electrons means that it
overcorrects band energies in materials severely. The first attempt to improve
this behavior, called sLOSC \cite{mahler2022b}, attenuated $J_{\bR ij}^\sigma$
empirically using a complementary error function, replacing the long-range $1/r$
behavior of the bare Hartree interaction by the exponentially decaying
$\erfc(\alpha r)/r$. The matrix elements of the sLOSC curvature are
\begin{equation}
\begin{split}
    \kappa_{\bR ij}^{\sigma}(\alpha, \beta)
        &= J_{\bR ij}^\sigma(\alpha) + \beta X_{\bR ij}^\sigma \\
        &= \iint d\br\, d\br'\, 
             \frac{\rho_{\bZ i \sigma}(\br) \erfc(\alpha \abs{\br - \br'})
             \rho_{\bR j \sigma}(\br')}
                 {\abs{\br - \br'}}
           + \beta \int d\br\, C_X \left[ \rho_{\bZ i \sigma}(\br) 
             \rho_{\bR j \sigma}(\br) \right]^{2/3},
\end{split}
\end{equation}
with the same $\beta, C_X$ as above; $\alpha = \SI{0.15}{\angstrom^{-1}}$ was
fit from a dataset of semiconductors and insulators.

However, the sLOSC curvature cannot be decomposed monochromatically because
the exchange term $X_{\bR ij}^\sigma$ is not linear in
$\ket{\rho_{\bR i \sigma}}$. In order to apply the monochromatic
decomposition to the sLOSC curvature, we replace the Dirac-type exchange $X$
by the exchange-correlation kernel
$f_{\xc}^{\sigma \sigma} = 
\delta^2 E_{\xc} / \delta \rho^\sigma \delta\rho^\sigma$, obtaining the modified
sLOSC curvature
\begin{equation}
    \kappa_{\bR ij}^\sigma(\alpha, \beta)
        = \mel{\rho_{\bZ i \sigma}}
              {\frac{\erfc(\alpha \abs{\br - \br'})}
                    {\abs{\br - \br'}} + 
                \beta f_{\xc}^{\sigma \sigma}}
              {\rho_{\bR j \sigma}}
        = \mel{\rho_{\bZ i \sigma}}{\wt{f}_{\Hxc}^{\sigma\sigma}(\alpha,\beta)}
              {\rho_{\bR j \sigma}}.
\end{equation}
This expression can be decomposed monochromatically, as the $\erfc$-modified
Coulomb kernel and the xc kernel scaled by $\beta$ remain linear in the DLWF
densities. The parameters $\alpha$ and $\beta$ must be reset empirically, which is
beyond the scope of this work, but the (modified) sLOSC curvature is thus
unified with the lrLOSC curvature. The implementation of the sLOSC curvature is
very simple: modify $f_{\Hxc}$ to include the attenuated Coulomb and scaled xc
interactions, and set $\ket{\delta \rho_{\bZ i \tau}} = 0$.

Finally, we note that the curvature of sLOSC (and of its predecessor, LOSC2
\cite{su2020}) are modified by a smoothing error function,
\begin{equation}
    \tilde{\kappa}_{\bR ij}^\sigma(\alpha, \beta) 
        = \erf(8 S_{\bR ij}^\sigma) \sqrt{\kappa_{ii\bZ}^\sigma
            \kappa_{jj\bZ}^\sigma} + 
           \erfc(8 S_{\bR ij}^\sigma) \kappa_{\bR ij}^\sigma,
\end{equation}
where $S_{\bR ij}^\sigma$ is the overlap
\begin{equation}
    S_{\bR ij}^\sigma = \int d\br\, 
        \left[\rho_{\bZ i \sigma}(\br) \rho_{\bR j \sigma}(\br)\right]^{1/2}.
\end{equation}
$S_{\bR ij}^\sigma$ is also nonlinear in $\ket{\rho_{\bR i \sigma}}$; to apply the
curvature smoothing monochromatically, we could use instead the modified overlap
\begin{equation}
    \widetilde{S}_{\bR ij}^\sigma = \left[ \int d\br\, 
        \rho_{\bZ i \sigma}(\br) \rho_{\bR j \sigma}(\br) \right]^{1/2}.
\end{equation}

\section{The lrLOSC Hamiltonian}
We provide more detail for the derivation of the correction provided by lrLOSC
to the DFA Hamiltonian. We have, applying the chain rule and the Wirtinger
derivative \cite{wirtinger1927}, that
\begin{equation}
    \Delta h^\sigma = 
    \frac{\delta \Delta E^\sigma}{\delta \rho^\sigma} =
    \sum_{ij\bR} 
        \left[ 
            \frac{\partial \Delta E^\sigma}{\partial \lambda_{\bR ij \sigma}}
                \frac{\delta \lambda_{\bR ij \sigma}}{\delta \rho^\sigma} +
            \frac{\partial \Delta E^\sigma}
                 {\partial \conj{\lambda}_{\bR ij \sigma}}
                \frac{\delta \conj{\lambda}_{\bR ij \sigma}}{\delta \rho^\sigma}
      \right].
\end{equation}
Recalling that
\begin{equation}
    \lambda_{\bR ij \sigma} = 
    \mel{w_{\bZ i \sigma}}{\rho^\sigma}{w_{\bR j \sigma}},
\end{equation}
we have by a matrix calculus identity that
\begin{equation}
    \frac{\delta \lambda_{\bR ij \sigma}}{\delta \rho^\sigma} =
    \frac{\delta \mel{w_{\bZ i \sigma}}{\rho^\sigma}{w_{\bR j \sigma}}}
         {\delta \rho^\sigma} =
    \ketbra{w_{\bR j \sigma}}{w_{\bR i \sigma}};
\end{equation}
similarly, 
\begin{equation}
    \frac{\delta \conj{\lambda}_{\bR ij \sigma}}{\delta \rho^\sigma} =
    \frac{\delta \mel{w_{\bR j \sigma}}{\rho^\sigma}{w_{\bZ i \sigma}}}
         {\delta \rho^\sigma} =
    \ketbra{w_{\bZ i \sigma}}{w_{\bR j \sigma}},
\end{equation}
since the complex conjugate of a matrix element of $\lambda$ is equal to the
corresponding element of its transpose.

Next, we observe that
\begin{equation}
    \frac{\partial \Delta E^\sigma}{\partial \lambda_{\bR ij \sigma}} =
    \frac{\partial
            \left[ 
                \frac12 \conj{\lambda}_{\bR ij \sigma}
                    \left( 
                        \delta_{\bR ij} - \lambda_{\bR ij \sigma}
                    \right) 
                \kappa_{\bR ij \sigma}
            \right]}
         {\partial \lambda_{\bR ij \sigma}} =
    -\frac12 \lambda_{ij\bR}^* \kappa_{ij\bR},
\end{equation}
and similarly
\begin{equation}
    \frac{\partial \Delta E^\sigma}{\partial \conj{\lambda}_{\bR ij \sigma}} =
    \frac{\partial 
        \left[ 
            \frac12 \conj{\lambda}_{\bR ij \sigma}
                \left(
                    \delta_{\bR ij \sigma} - \lambda_{\bR ij \sigma}
                \right) 
            \kappa_{\bR ij \sigma}
        \right]}
         {\partial \conj{\lambda}_{\bR ij \sigma}} =
    \frac12 
        \left( 
            \delta_{\bR ij \sigma} - \lambda_{\bR ij \sigma}
        \right)
        \kappa_{\bR ij \sigma}.
\end{equation}
Combining these terms, we obtain that
\begin{equation}
\begin{split}
    \Delta h^\sigma &=
    \sum_{\bR ij} \frac12 \kappa_{\bR ij \sigma}
        \left[
            \left( \delta_{\bR ij} - \lambda_{\bR ij \sigma} \right)
                \ketbra{w_{\bZ i \sigma}}{w_{\bR j \sigma}} -
            \conj{\lambda}_{\bR ij \sigma} 
                \ketbra{w_{\bR j \sigma}}{w_{\bZ i \sigma}}
        \right] \\ &=
    \sum_{\bR ij} \Re{\kappa_{\bR ij \sigma}} 
        \left( \frac12 \delta_{\bR ij} - \lambda_{\bR ij \sigma} \right)
            \ketbra{w_{\bZ i \sigma}}{w_{\bR j \sigma}}.
\end{split}
\end{equation}

\section{Note on symmetry and degeneracy}
\begin{figure*}[ht]
    \centering
    \includegraphics[width=\textwidth]{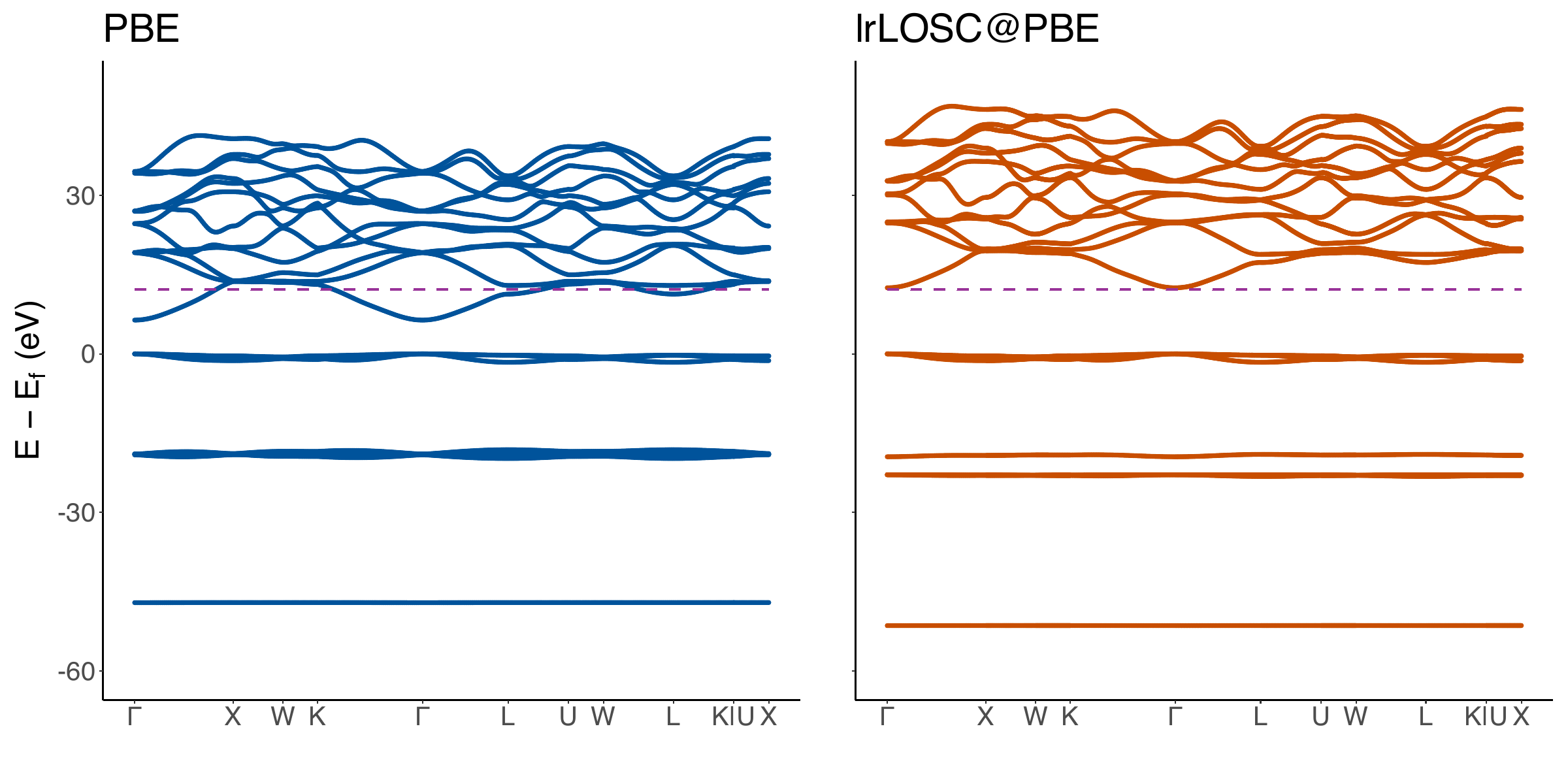}
    \caption{NaF band structure. Purple dashed line: ZPR-corrected experimental
             gap.}
    \label{fig:naf}
\end{figure*}

lrLOSC usually shifts the energy each isolated group of bands by an amount
nearly constant across the Brillouin zone. Occupied states are usually shifted
down, and virtual states up; core states may be shifted more than valence ones. 
The overall shape of the DFA band structure, as well as the valence and
low-lying conduction states' degeneracies, are \emph{generally} preserved. We
have observed a few instances of degeneracy breaking, however. The most obvious
case is NaF, in which lrLOSC opens a small gap among the four shallow core
states (FIG.~\ref{fig:naf}).

\begin{figure*}
    \centering
    \includegraphics[width=\textwidth]{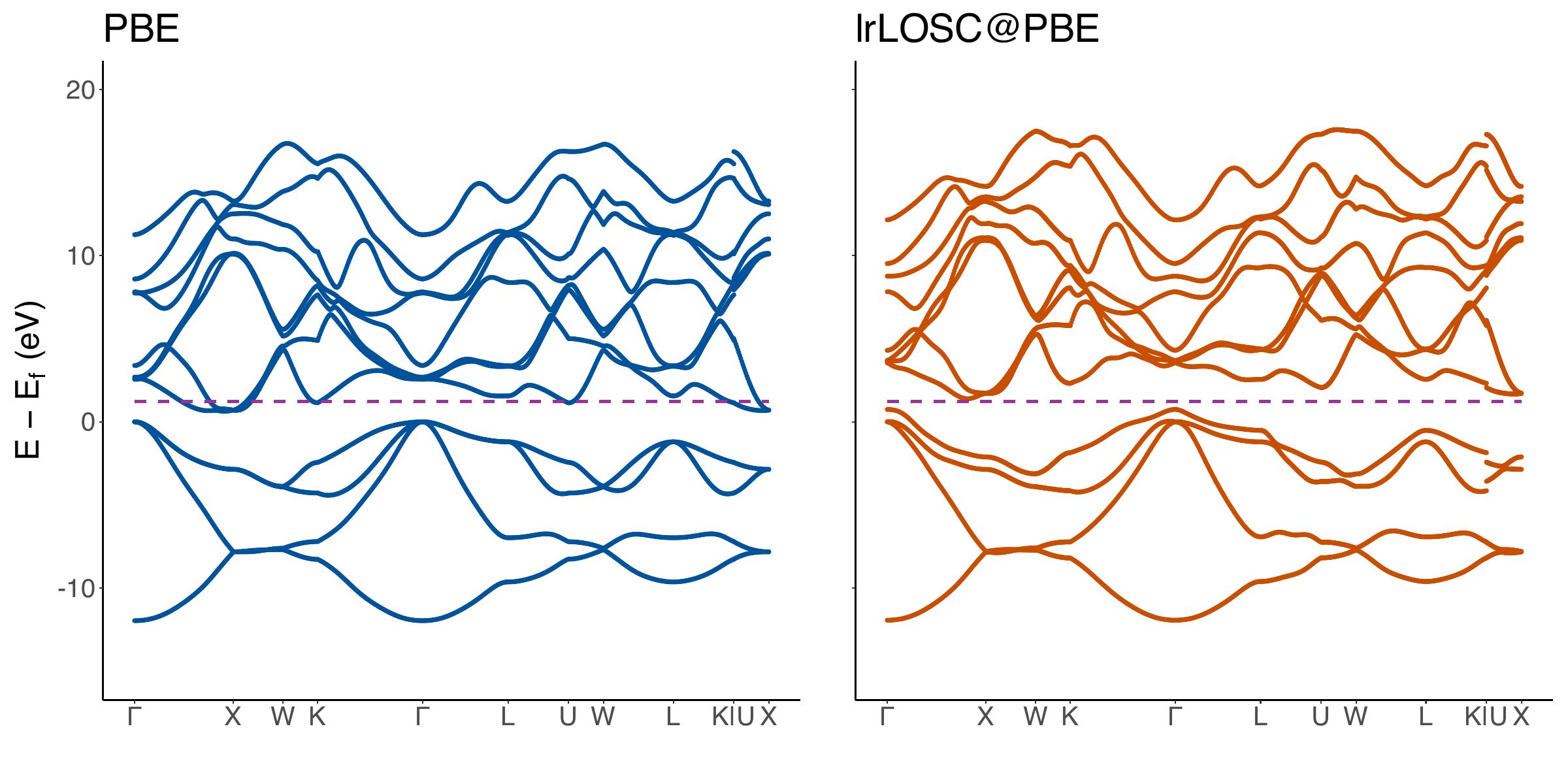}
    \caption{Si band structure. Purple dashed line: ZPR-corrected experimental
             gap.}
    \label{fig:si}
\end{figure*}

As seen in FIG.~\ref{fig:si}, lrLOSC also breaks the threefold degeneracy in
the highest occupied bands of silicon at $\Gamma$. These should be protected by
symmetry; they transform together as $\Gamma_{25'}$ in the diamond lattice
\cite{buberman1971}. We do not enforce these symmetries in either the DLWFs or
the computation of the curvature (although the PBE band structure is
interpolated from the DLWFs, so it is the linear-response curvature that breaks
the degeneracy \cite{gonze1995b}). We could address this degeneracy breaking by,
e.g., implementing degenerate density functional perturbation theory
\cite{palenik2016a, palenik2017a}; as the impact of the degeneracy breaking is
fairly small, we leave this to future study.

\section{Verification of the unit cell-periodic implementation}
The previous implementation of LOSC for materials \cite{mahler2022b} did not
include the monochromatic decomposition; its curvature elements $\kappa_{\bR ij}$
were computed on the large supercell. 

We compared the self-Hartree energy
\begin{equation}
    J_{ii\bZ}^\sigma = \iint d\br\, d\br'\, \rho_{\bZ i \sigma}(\br)
        K(\abs{\br - \br'}) \rho_{\bZ i \sigma}(\br')
\end{equation}
of the DLWFs corresponding to the HOMO and LUMO of water and silicon across
the implementations. $K(r)$ is the Hartree kernel, which can be a bare Coulomb
interaction $K(r) = 1/r$ or (for sLOSC) the attenuated
$K(r) = \erfc(\alpha r)/r$. The Coulombic divergence in reciprocal space is
handled with various corrections: the spherical cutoff (sphcut)
\cite{rozzi2006}, Gygi--Baldereschi (G--B) \cite{gygi1986} for $\bk$-sampled
systems like Si), or Martyna--Tuckerman (M--T) \cite{martyna1999} for isolated
systems like water. All computations use the PBE functional \cite{perdew1996},
and the $J_{ii\bZ}$ values are in rydbergs.

We simulated a single water molecule in a \SI{25}{\angstrom} cubic cell, with a
kinetic energy cutoff of \SI{100}{\rydberg}; to emulate an isolated system, it
is sampled only at the origin $\Gamma$ of the Brillouin zone. We compare it
against the in-house \texttt{QM$^4$D} code, using the aug-cc-pvTZ basis set.
In the tabes below, \texttt{QE-SC} refers to the original, supercell-periodic
implementation of LOSC, and \texttt{QE-UC} refers to the monochromatically
decomposed implementation.
\begin{table}[ht]
\centering
\begin{tabular}{@{}rrrrr@{}}
    \toprule
    Code             & Kernel & Correction  & $J_{\homo}$ & $J_{\lumo}$ \\
    \midrule
    \texttt{QM$^4$D} & bare   & n/a         & 1.436 & 0.413 \\
    \texttt{QE-SC}   & bare   & sphcut      & 1.442 & 0.415 \\ 
    \texttt{QE-UC}   & bare   & M--T        & 1.442 & 0.415 \\
    \texttt{QE-UC}   & bare   & sphcut      & 1.442 & 0.415 \\
    \texttt{QM$^4$D} & erfc   & n/a         & 1.110 & 0.155 \\
    \texttt{QE-SC}   & erfc   & sphcut      & 1.115 & 0.157 \\
    \texttt{QE-UC}   & erfc   & none        & 1.115 & 0.157 \\
    \bottomrule
\end{tabular}
\caption{Self-Hartree energy for the DLWFs correponding to the HOMO and LUMO of water.}
\label{tab:waterkap}
\end{table}

We additionally simulated silicon, with the same unit cell as in
Table~\ref{tab:edata} below; here, however, we used a $4 \times 4 \times 4$
Monkhorst--Pack mesh centered at $\Gamma$, with a \SI{100}{\rydberg} kinetic
energy cutoff.
\begin{table}[ht]
\centering
\begin{tabular}{@{}rrrrr@{}}
    \toprule
    Code & Kernel & Divergence & $J_{\vbm}$ & $J_{\cbm}$ \\
    \midrule
    \texttt{QE-SC}   & bare   & sphcut      & 0.525 & 0.473 \\ 
    \texttt{QE-UC}   & bare   & G--B        & 0.555 & 0.510 \\
    \texttt{QE-UC}   & bare   & sphcut      & 0.525 & 0.473 \\
    \texttt{QE-SC}   & erfc   & sphcut      & 0.258 & 0.213 \\
    \texttt{QE-UC}   & erfc   & none        & 0.258 & 0.213 \\
    \bottomrule
\end{tabular}
\caption{Self-Hartree energy (\si{\rydberg}) for DLWFs corresponding to the
         valence band maximum (VBM) and conduction band minimum (CBM) of
         silicon.}
\label{tab:sikap}
\end{table}

The plane-wave codes \texttt{QE-SC} and \texttt{QE-UC} show good agreement with
the Gaussian-type orbital code; both the Martyna--Tuckerman and spherical cutoff
correction in the monochromatic implementation agree precisely with the
spherical cutoff in the supercell-periodic implementation. In
Table~\ref{tab:sikap}, we can see that the spherical cutoff is perfectly
replicated in the monochromatic decomposition compared to the
supercell-periodic one; however, the Gygi--Baldereschi correction yields a
different (and larger) Coulomb repulsion than does the spherical cutoff. The
methods are not equivalent; this indicates that a larger $\bk$-mesh is required
to use the spherical cutoff, since with only a $4 \times 4 \times 4$ mesh some
of the Coulomb tail is being removed.

Note additionally that the \texttt{QE-SC} implementation can use a the
spherical cutoff in addition to the $\erfc$-attenuated kernel; \texttt{QE-UC}
cannot combine the two, but the exponential decay is fast enough that the
results are numerically equal in the systems tested.

\section{Data table}
Table~\ref{tab:edata} below contains structural parameters, total energy, and
band gaps from PBE, lrLOSC, and (where applicable) experiment. The lattice
constants $a$ were taken from \citet{heyd2005} and the references therein
whenever available, and from \citet{wyckoff1973} otherwise. Experimental band
gaps are from \citet{tran2009} (LiF, LiCl, Ne, Ar); \citet{poole1975} (NaF);
\citet[p.~184]{madelung2004} (MgO); and \citet{heyd2005} and containing
references (C, Si, SiC, Ge, BN, BP, AlP).
Zero-point renormalization values are computed by \citet{engel2022},
(C, Si, SiC, LiF, MgO, BN, AlP), \citet{shang2021} (BP, LiCl, NaF), and
\citet{miglio2020} (Ge).

We exclude Ne and Ar from Figure 1 in the main text and from the calculation of
the band gap MAE because they do not yet have ZPR calculations available.
Observe that lrLOSC underestimates their experimental band gaps, although the
result is still much better than PBE's; if the ZPR is negative, as is the case
for the rest of the systems, the electronic gap will be closer to lrLOSC's
prediction than shown in Fig.~1 of the main text.

\begin{table}[htb]
\centering
\caption{Detailed structural information, total energies, and band gaps.
Lattice constants are from experiments (see above).
ZPR is calculated theoretically \cite{engel2022, shang2021, miglio2020}.
Error: lrLOSC $-$ electronic gap.}
\label{tab:edata}
\setlength\tabcolsep{0pt}
\begin{tabular*}{\linewidth}{@{\extracolsep{\fill}}rrrrrrrrrrr}
\toprule
\multicolumn{3}{c}{System parameters} & 
\multicolumn{2}{c}{Total energy (\si{\rydberg})} &
\multicolumn{6}{c}{Band gaps (\si{\electronvolt})} \\
\cmidrule(lr){1-3} \cmidrule(lr){4-5} \cmidrule(lr){6-11}
Name & Structure & $a$ (\si{\angstrom}) & 
$E_{\text{PBE}}$ & $\Delta E_{\LOSC}$ &
PBE & lrLOSC & Exp. & ZPR & Elec. & Error \\
\midrule
C & diamond & 3.567 & $-$24.075 & 2.10$\times 10^{-5}$ & 4.205 & 5.674 & 5.48 & $-$0.323 & 5.803 & $-$0.13 \\
Si & diamond & 5.43 & $-$16.923 & 1.19$\times 10^{-3}$ & 0.705 & 1.711 & 1.17 & $-$0.058 & 1.228 & 0.48 \\
SiC & zincblende & 4.358 & $-$20.539 & 1.72$\times 10^{-4}$ & 1.356 & 2.635 & 2.42 & $-$0.175 & 2.595 & 0.04 \\
Ge & diamond & 5.658 & $-$357.28 & 1.87$\times 10^{-3}$ & 0.032 & 0.798 & 0.74 & $-$0.033 & 0.773 & 0.02 \\
MgO & rocksalt & 4.207 & $-$152.10 & 5.94$\times 10^{-5}$ & 4.803 & 8.727 & 7.97 & $-$0.533 & 8.503 & 0.22 \\
LiF & rocksalt & 4.017 & $-$63.982 & 1.89$\times 10^{-5}$ & 9.187 & 15.24 & 14.20 & $-$1.231 & 15.431 & $-$0.19 \\
LiCl & rocksalt & 5.13 & $-$47.474 & 1.34$\times 10^{-4}$ & 6.329 & 9.783 & 9.40 & $-$0.436 & 9.836 & $-$0.05 \\
NaF & rocksalt & 4.62 & $-$140.97 & 4.59$\times 10^{-5}$ & 6.39 & 12.502 & 11.50 & $-$0.699 & 12.199 & 0.30 \\
BN & zincblende & 3.616 & $-$26.805 & 1.88$\times 10^{-5}$ & 4.531 & 6.387 & 6.22 & $-$0.402 & 6.622 & $-$0.23 \\
BP & zincblende & 4.538 & $-$19.792 & 2.60$\times 10^{-4}$ & 1.345 & 2.769 & 2.40 & $-$0.101 & 2.501 & 0.27 \\
AlP & zincblende & 5.463 & $-$18.576 & 8.40$\times 10^{-4}$ & 1.576 & 3.026 & 2.51 & $-$0.096 & 2.606 & 0.42 \\
Ne & cubic & 4.429 & $-$66.715 & 6.21$\times 10^{-6}$ & 11.616 & 19.392 & 21.70 & --- & --- & --- \\
Ar & cubic & 5.256 & $-$45.18 & 3.96$\times 10^{-5}$ & 8.703 & 14.028 & 14.20 & --- & --- & --- \\
\midrule
\multicolumn{10}{r}{MAE (\si{\electronvolt})} & 0.22 \\
\bottomrule
\end{tabular*}
\end{table}

\section{Comparison of lrLOSC with other methods}
In Table~\ref{tab:method-compare}, we compare the band gaps obtained from lrLOSC
with those computed by $G_0W_0$ \cite{chen2015a, shishkin2007a}, qs$GW$
\cite{chen2015a}, equation-of-motion coupled cluster with singles and doubles
(CCSD), \cite{vo2024},  Koopmans-compliant integral (KI) functionals
\cite{nguyen2018, colonna2022}, the Wannier--Koopmans (W--K) method
\cite{ma2016, weng2017}, and the Wannier optimally tuned screened
range-separated hybrid method (SRSH) \cite{wing2021}.
Only systems represented by every method---in this case, \ce{C}, \ce{Si},
\ce{MgO}, and \ce{AlP}---are used to compute the mean absolute error (MAE), so
those presented in this table differ from those in the main text of their
respective papers. Despite the small dataset, lrLOSC's performance is
competitive with the other methods; its maximum deviation is smaller than
any method except SRSH, and its MAE is within \qty{0.2}{\electronvolt} of the
best methods (for these four materials, CCSD and SRSH).

This comparison is not, of course, completely rigorous. We show systems computed
by lrLOSC in addition to the other methods; each method additionally computes
the band gap for systems we did not consider, and those are not shown here. We
have not ensured that the lattice constants and underlying DFA parameters are
identical, and not all methods treat spin-orbit coupling, e.g., in \ce{Ge},
equally. Nevertheless, it may be useful as a rough reference to compare the
different methods.

\begin{table}[htb]
\centering
\caption{Electronic band gaps (in \si{\electronvolt}) by different methods.
Experimental gaps are adjusted for zero-point renormalization (ZPR).}
\label{tab:method-compare}
\begin{tabular*}{\linewidth}{@{\extracolsep{\fill}}rrrrrrrrrr}
\toprule
System & $\text{Exp.} - \text{ZPR}$ & 
PBE & lrLOSC & $G_0W_0$ & qs$GW$ & CCSD & KI   & W--K & SRSH \\
\midrule
C & 5.80 & 4.20 & 5.67 & 5.59 & 6.40 & 5.85 & 6.30 & 5.82 & 5.70 \\
Si & 1.23 & 0.70 & 1.71 & 1.17 & 1.47 & 0.96 & 1.22 & 1.11 & 1.10 \\
MgO & 8.50 & 4.80 & 8.73 & 7.89 & 9.29 & 8.34 & 7.61 & 8.00 & 8.20 \\
AlP & 2.60 & 1.58 & 3.03 & 2.49 & 3.10 & 2.62 & 2.63 & 2.44 & 2.60 \\
\midrule
MAE & --- & 1.71 & 0.32 & 0.25 & 0.53 & 0.13 & 0.36 & 0.20 & 0.13 \\
\midrule
SiC & 2.60 & 1.36 & 2.64 & 2.25 & 2.90 & 2.54 & 2.48 & 2.69 & --- \\
Ge & 0.77 & 0.03 & 0.80 & 0.50 & 0.96 & --- & 1.00 & 0.63 & 0.70 \\
LiF & 15.40 & 9.19 & 15.20 & 13.30 & --- & 15.40 & 13.90 & 14.30 & 15.40 \\
LiCl & 9.84 & 6.34 & 9.78 & 9.27 & 11.00 & 9.43 & --- & 9.68 & --- \\
NaF & 12.20 & 6.39 & 12.50 & --- & --- & --- & --- & 11.20 & --- \\
BN & 6.62 & 4.53 & 6.39 & 6.19 & --- & 6.45 & 7.04 & --- & --- \\
BP & 2.50 & 1.34 & 2.77 & --- & --- & 1.65 & 2.30 & --- & --- \\
\midrule
Maximum deviation & --- & 6.21 & 0.48 & 2.1 & 1.16 & 0.85 & 1.5 & 1.1 & 0.3 \\
\bottomrule
\end{tabular*}
\end{table}

\bibliography{lrLOSC}